\documentclass[11pt]{article} \usepackage{moriond,epsfig}

\def\Journal#1#2#3#4{{#1} {\bf #2}, #3 (#4)}

 \def\AA{\em A\&A} \def\APJ{\em ApJ}
\def\SCI{\em Science}

\def\be{\begin{equation}} \def\ee{\end{equation}}
\def\bea{\begin{eqnarray}} \def\eea{\end{eqnarray}}

\def\HESS{H.E.S.S.} 
\begin{document} \vspace*{4cm} \title{Very High Energy $\gamma$-ray
Observations with \HESS}

\author{Christopher van Eldik for the \HESS\ Collaboration}

\address{Max-Planck-Institut f\"ur Kernphysik, P.O. Box 103980,
D-69029 Heidelberg, Germany}

\maketitle\abstracts{ The \HESS\  Imaging Atmospheric Cherenkov
  Telescope Array 
is currently the most sensitive instrument for Very High Energy (VHE)
$\gamma$-ray observations in the energy range of about
0.1-10~TeV. During more than two years of operation with the complete
4-telescope array, many galactic and extragalactic VHE $\gamma$-ray
sources have been discovered. With its superior sensitivity and its
large field-of-view camera, \HESS\  is particularly suited for surveys
and detailed studies of extended sources. A selection of recent
\HESS\  results is presented in this proceeding.  }

\section{The \HESS\  Experiment} \HESS\  is an array of four Imaging
Atmospheric-Cherenkov Telescopes (IACTs) located in the Khomas Highlands of
Namibia (S 23$^\circ$16'18'' E 16$^\circ$30'1'') at an altitude of
1800~m above sea level. Each telescope consists of a segmented
107~m$^2$ mirror of 13~m diameter which reflects Cherenkov photons
onto a 960~pixel photomultiplier camera \cite{Camera} with a large
field-of-view of about $5^\circ$. The focal length of the instrument
15~m.\\ \HESS\  makes use of the stereoscopic technique, i.e. air
showers are imaged by several telescopes simultaneously. In order to
suppress background (mostly due to muons) at the hardware level
and to enable stereoscopic reconstruction, a stereo trigger
\cite{Trigger} accepts only those events which were seen by at least
two telescopes.  The reconstruction of the air shower from different
points of view leads to a superior angular resolution of better than
$0.1^\circ$ per event. Positions of strong point-like sources are
determined with a precision of 20-30'', limited by the
pointing accuracy of the array. Studies to further reduce the
pointing systematics are ongoing.  The energy of the primary
$\gamma$-ray is estimated with a typical energy resolution of less
than 15\%. The energy threshold of the array before analysis cuts is
about 100~GeV for observation at zenith and increases to about 700~GeV
at $30^\circ$ of altitude.\\ 
\HESS\  is currently the most sensitive
instrument in its field: a significant ($5~\sigma$) detection of
point-like sources with a Crab-like flux takes 30~seconds. Fainter
sources with fluxes of only 1\% of the Crab Nebula are detected in
25~hours of observation time.  \HESS\ reaches a good off-axis performance in its large field-of-view cameras. This makes the instrument
ideal for surveys and studies of extended sources.\\ 
A first-look data analysis is performed during data taking. This
enables one to monitor the data quality and quickly react to any
malfunctions of the system. Furthermore, re-observations can be
scheduled immediately in case a particular target shows interesting
physics. Final calibration and analysis of the data is later performed in
Europe, where two independent calibration and several analysis chains are used 
to cross-check results. For more information on analysis techniques
see\cite{Analysis}.

\section{Selected results from \HESS\ } \HESS\  recorded useful data
during comissioning of the array (with a reduced number of
telescopes and a number of different trigger configurations). The
instrument was completed in December 2003 and is operating since then
with full performance. \HESS\  observations have resulted in far too numerous
results than could be presented here. Therefore, for the purpose of
this paper, only a selection of recent \HESS\ highlights will be presented.

\subsection{Galactic Plane survey} 
\begin{figure}[htpb]
  \begin{center}
  \includegraphics[width=\textwidth]{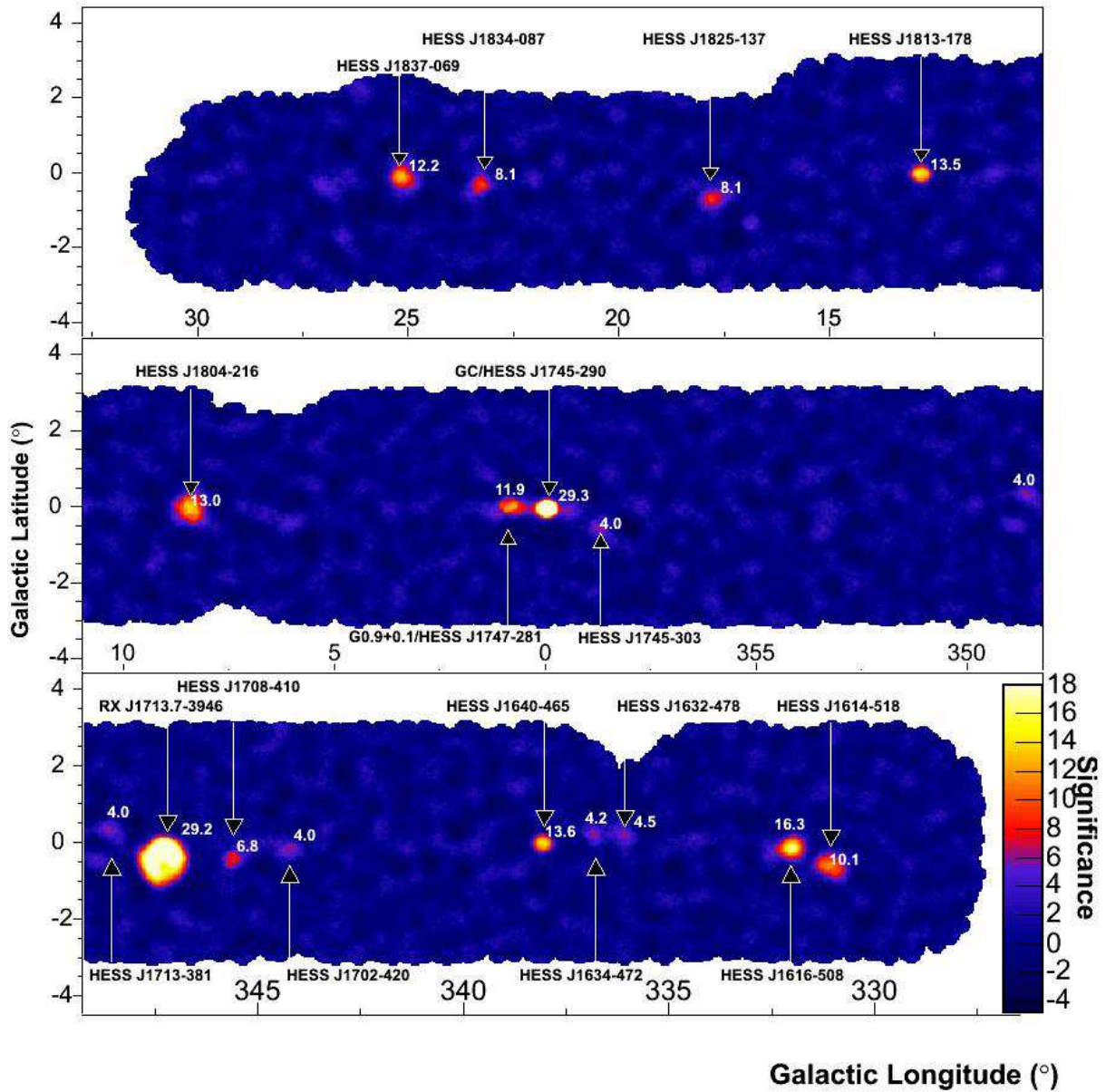}
  \end{center}
  \caption{Significance map of the \HESS\  Galactic Plane survey in
    2004, including data from re-observations of source candidates from
    the original scan and observations of the known $\gamma$-ray sources
    RX~J1713.7-3946 and the Galactic Centre region.}
  \label{fig:Scan}
\end{figure}

The location of \HESS\  in the
southern hemisphere makes the instrument ideal for observations of the
Galactic Plane, in particular in the direction of the Galactic
centre. The inner part of the Galactic Disk is expected to host a rich
population of all types of potential multi-TeV particle accelerators
like pulsars, supernova remnants (SNRs), and massive stars. In the
central $60^\circ$ in galactic longitude and $6^\circ$ in galactic
latitude, 91 SNRs and 381 pulsars have so far been detected in
different energy bands \cite{Green04,Man05}.\\ 
Prior to the start of \HESS\ observations, only two strong Galactic
sources were known in VHE 
$\gamma$-rays: TeV $\gamma$-ray emission from the Galactic Centre had
been reported by the Whipple \cite{whip04} and Cangaroo \cite{Cang04}
collaborations. The latter also claimed TeV emission from the SNR RX
J1713.7-3946 \cite{Cang02}, later confirmed by \HESS\  measurements in
2003 \cite{rxj03}.\\ 
The potentially large number of $\gamma$-ray sources combined with the
enhanced sensitivity of \HESS\  motivated a deep scan of the inner part
of the Galactic Plane \cite{ScanI,ScanII}, which was carried out in
2004. The survey covered a range of $-30^\circ < l < 30^\circ$ and
$-3^\circ < b < 3^\circ$ in Galactic longitude and latitude,
respectively. In 230 hours of live time, an average flux sensitivity
of about 2\% of the Crab above 200~GeV was reached. The scan revealed
fourteen previously unknown VHE $\gamma$-ray sources with a
significance $>4~\sigma$ (Fig. \ref{fig:Scan}), all of which are
extended in shape when accounting for the angular resolution of the
instrument. Due to good pointing accuracy most of the sources could be
identified with already known counterparts in other wavelength
bands. There is, however, a substantial fraction of sources for which
there exists no known counterpart yet. Whether or not these are driven
by yet undiscovered physics processes (e.g. Dark Matter annihilation)
remains one of the exciting tasks for future investigations.\\
Integral fluxes and energy spectra have been
obtained for all sources. Energy spectra can be described
by power-law parametrisations
\begin{equation}
  \label{eq:one} \frac{dN}{dE} = F_0 E^{-\Gamma},
\end{equation} and similar values of the photon index $\Gamma$ were
found for all sources. The average value of $\Gamma=2.32$ matches
expectations for the source spectrum of Galactic cosmic rays
\cite{Jon2001}. Recently, the MAGIC collaboration has confirmed two of
the newly discovered \HESS\  sources, HESS J1813-178 \cite{MAGIC1} and
HESS J1834-087 \cite{MAGIC2}. Their integral fluxes and photon indices
agree well with those from \HESS\  observations.\\ 

\subsection{Supernova remnants} 
\begin{figure}[htpb]
  \begin{center}
    \includegraphics[width=0.7\textwidth]{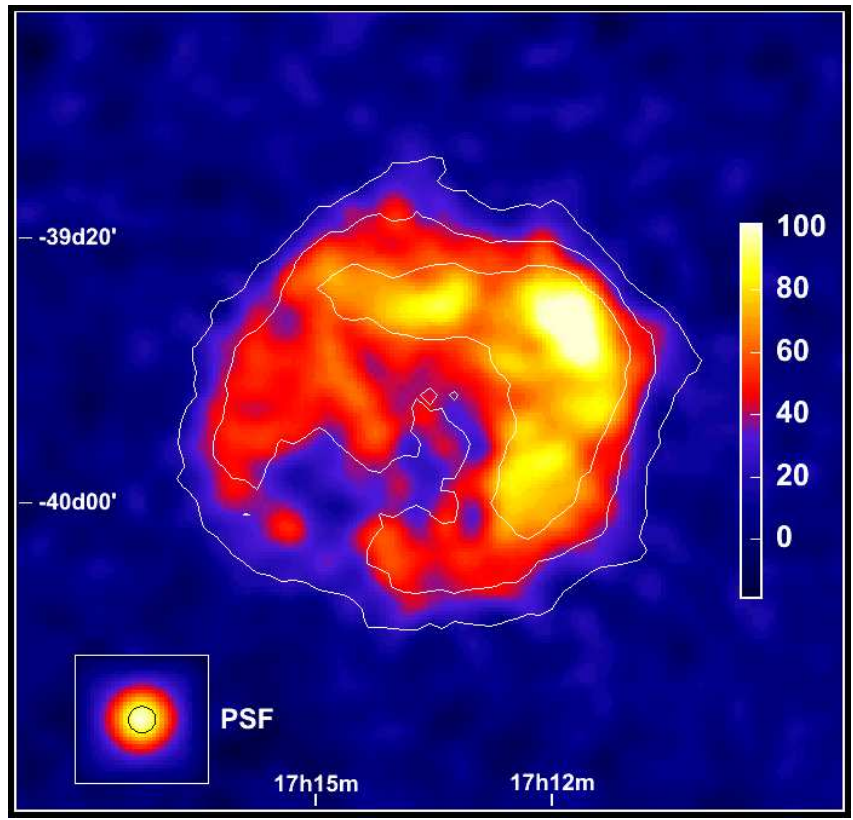}
    \caption{Image of RXJ 1713.7-3946 in VHE $\gamma$-rays. The linear
      colour scale is in units of excess counts. The white contours denote
      the significance of the features. The levels are linearly spaced and
      correspond to 5, 10, and 15$\sigma$, respectively. In the lower left
      corner a simulated point source is shown as it would appear in this
      particular data set.}
  \label{fig:rxj1713}
  \end{center}
\end{figure}

Supernova remnants, left behind by
gigantic explosions of massive stars, have since long been
thought to be one of the prime sources of VHE cosmic rays. Shock
acceleration is believed to be the mechanism accelerating charged
particles to energies of 100~TeV and beyond, naturally resulting in a
power-law shaped energy spectrum. VHE $\gamma$-rays can trace those
particles by means of two common scenarios: If the accelerated
particels were electrons, they could produce $\gamma$-rays by means of
Inverse Compton scattering off ambient photons. Protons and nucleons
could undergo nucleonic reactions with ambient material and produce,
among other particles, neutral pions, followed by the decay
$\pi^{0}\to\gamma\gamma$.\\ At least 5 sources seen in the \HESS\ 
Galactic Plane scan could be identified as SNRs. One of the best
studied examples is RXJ~1713.7-3946, a shell-type supernova remnant,
which was observed by \HESS\  in 2003. With an angular
resolution more than an order of magnitude better than the spatial
extension of the SNR, \HESS\  was the first to resolve the morphology of
a TeV $\gamma$-ray source \cite{rxj03}. In 2004, RXJ 1713.7-3946 was
re-observed for 33 hours live time. A strong signal is observed with a
significance of about $39\sigma$ (Fig. \ref{fig:rxj1713}), which
enables detailed spectral and morphological studies \cite{rxj06}. Good
spatial correlation of the $\gamma$-ray flux with X-rays proves that
SNRs are indeed accelerators of VHE particles. Although the
interpretation of the \HESS\  results in the context of data in other
energy bands seems to favour a hadronic acceleration scenario, no
strong conclusion can yet be drawn about the parent population
dominantly responsible for the observed $\gamma$-ray flux.

\subsection{Galactic Centre region} The inner 100~parsec of our
Galaxy are known as the most violent and active region in our solar
neighbourhood. Not only does it host a central supermassive black
hole, but also numerous other objects like supernove remnants, massive
stars and giant molecular clouds.\\ 
In 2003 \HESS\  observed a strong source of TeV
\begin{figure}[htpb]
  \begin{center}
  \includegraphics[width=0.9\textwidth]{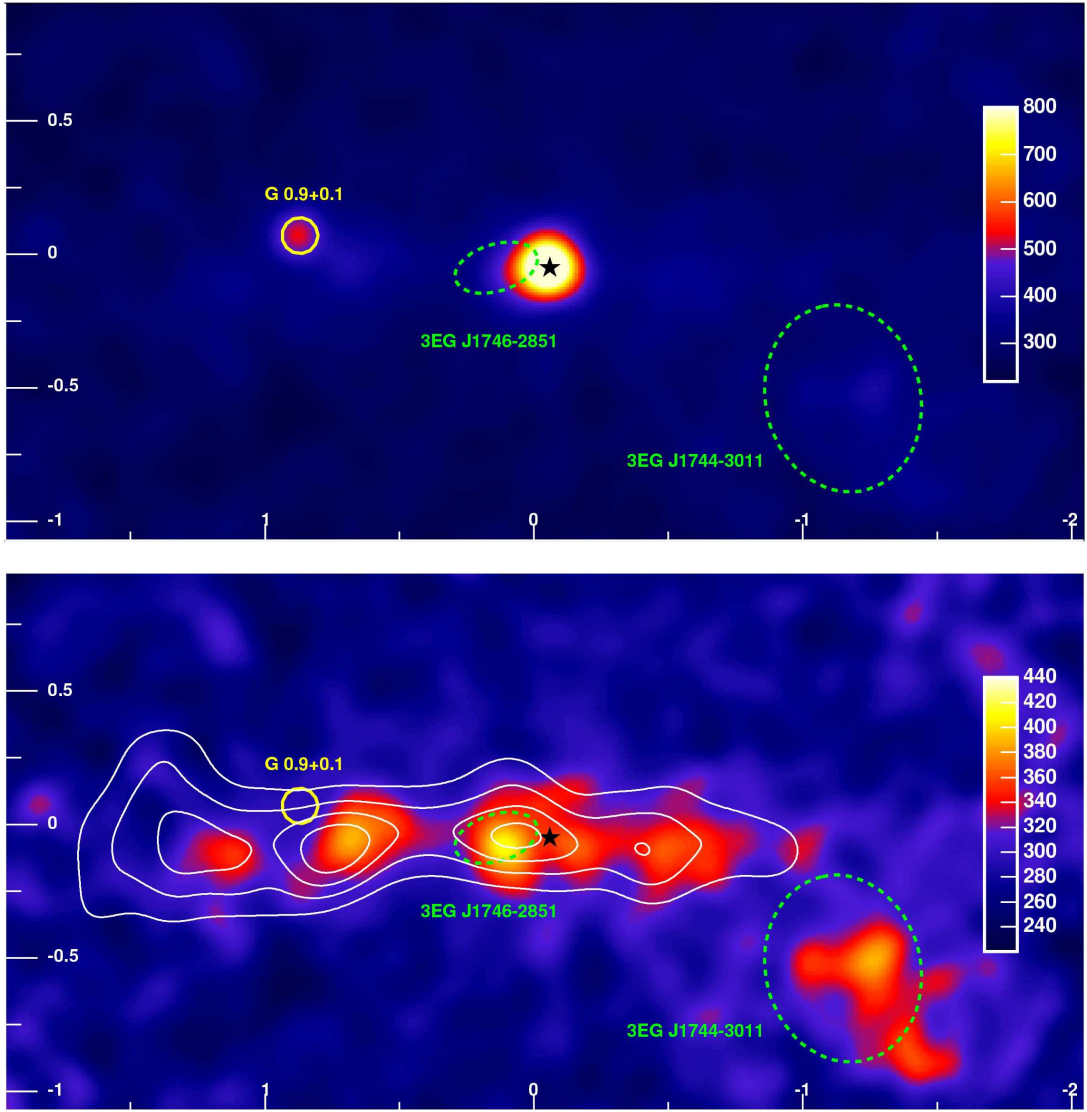}
  \caption{VHE $\gamma$-ray images of the GC region. (a) $\gamma$-ray
count map, showing the strong source G0.9+0.1 and the emission
coincident with the gravitational centre of our galaxy. The position
of Sagittarius A$^*$ is marked with a black star. (b) Same map after
subtraction of the two bright sources. White contour lines indicate
the density of dense molecular clouds, traced by CS emission. The
green ellipses show 95\% confidence regions for the position of two
unidentified EGRET sources.}
  \label{fig:GCDiffuse}
  \end{center}
\end{figure} 
$\gamma$-rays (HESS J1745-209) from the direction of the
Galactic Centre \cite{GC2003}. The Galactic Centre region was
re-observed in 2004, and the initial discovery was confirmed with a
high significance of about $38~\sigma$
(Fig. \ref{fig:GCDiffuse}~(a)). The centre of gravity of the (almost)
pointlike excess is spatially coincident ($3''\pm 12'' (stat.)$) with
the central black hole Sagittarius A$^*$. However, even with the good
pointing accuracy of the instrument (20''), the SNR
Sagittarius A-East can not be ruled out as the source of the
observed TeV $\gamma$-ray flux.\\ 
The energy spectrum of the excess is shown in
\begin{figure}[htpb]
  \begin{center}
  \includegraphics[width=\textwidth]{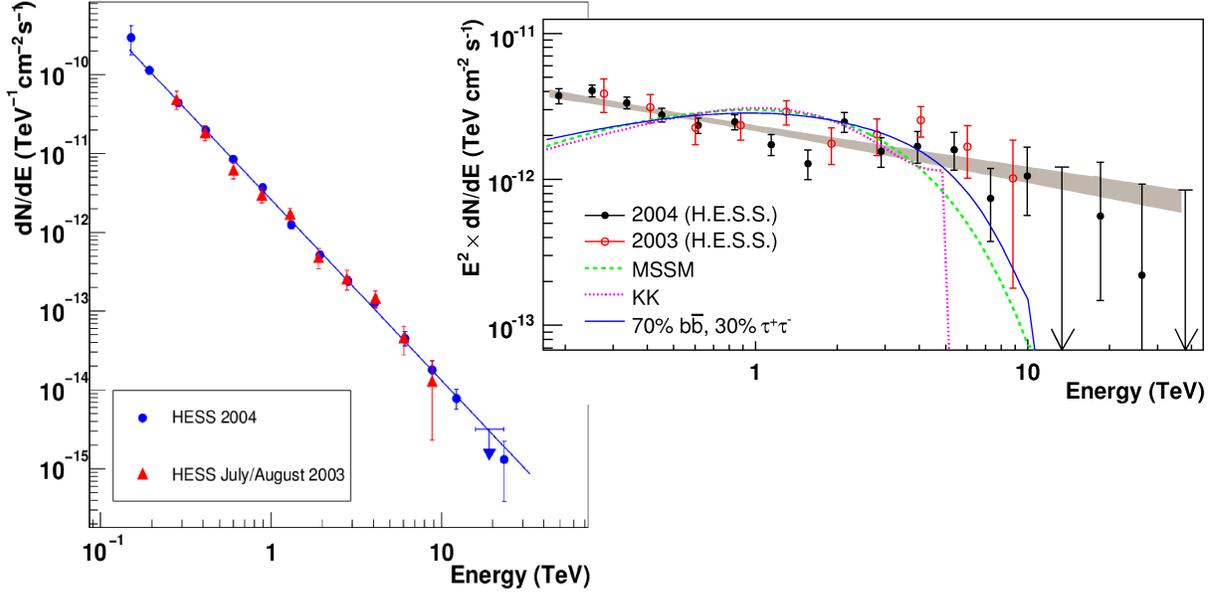}
  \caption{Energy spectrum of HESS J1745-209. Left: shown is the
differential $\gamma$-ray flux as a function of the reconstructed
energy for both the 2003 and 2004 data from observations of the
Galactic Centre. The solid curve shows a power-law fit to the 2004
data points. Right: spectral energy distribution including best-fit
curves of different predictions of dark matter annihilation
models. Clearly data rule out that the observed $\gamma$-ray flux is
due to dark matter annihilation only.}
  \label{fig:GCEnergy}
  \end{center}
\end{figure} 
Fig. \ref{fig:GCEnergy} (left side) separately for the
2003 and 2004 data sets. The 2004 data match the previous results both
in observed flux and photon index. The energy dependence is best
described by a straight power-law of index $\Gamma=2.25$, with no hint
for a break or curvature in the spectrum up to energies of 20~TeV. A
detailed analysis shows no variability of the VHE flux
or the photon index on time scales of years, months, days, hours,
or minutes. While a variable source would disfavour an SNR or, more
exotically, annihilation of Dark Matter particles, as the source of
the observed TeV $\gamma$-ray excess, no firm conclusion can yet be
drawn from the non-observation of such variability.\\ Annihilation of
light Dark Matter (DM) particles (neutralinos in most common DM
scenarios) in the vicinity of the central black hole Sagittarius A$^*$
can be searched for by comparing the predicted energy spectra of the
$\gamma$-rays produced in the annihilation process with the energy
spectrum measured by \HESS\ Fig. \ref{fig:GCEnergy} (right side)
shows the spectral energy density ($E^2\ \times$ differential flux) of
the observed excess together with best-fit curves of predictions of
different annihilation models.  As can be seen, all predicted
spectra are curved at high energies (for reasons of energy
conservation), clearly in contradiction to the observed power-law
behaviour of the 2004 \HESS\  measurements. Secondly, the neutralino
masses determined by the fits are rather large compared to common
expectations. As a consequence, the bulk of the observed $\gamma$-ray
excess is probably of astrophysical rather than of particle physics
origin. However, a small admixture of $\gamma$-rays from DM
annihilations cannot be ruled out.\\ While for previous VHE
instruments the sources shown in Fig.  \ref{fig:GCDiffuse}~(a) were
close to the detection limit, the sensitivity of \HESS\  enables the
search for much fainter emission.  Subtracting the best-fit model for
point-like emission of the two strong sources HESS~J1745-209 and
G~0.9+0.1 yields the map shown in Fig. \ref{fig:GCDiffuse}~(b). The
subtraction reveals the presence of diffuse emission along the
Galactic Plane \cite{GCDiffuse06}, as well a a TeV $\gamma$-ray
source \cite{ScanII} coincident with the unidentified EGRET source
3EGJ1744-3011.\\ The diffuse emission spans in a region of roughly
$2^\circ$ in galactic longitude with an rms width of about $0.2^\circ$
in galactic latitude. The reconstructed $\gamma$-ray spectrum
integrated within $|l|\leq 0.8$ and $|b|\leq 0.3$ is well described by
a power law with photon index $\Gamma=2.29$, close to what is found
for the bright central source.\\ Assuming the diffuse emission being
produced near the centre of the Galaxy (at a distance of 8.5~kpc from
the observer), the latitude extension translates into a scale of about
30~kpc. This is very similar in extension to giant molecular clouds in
this region. Indeed, at least for $|l|\leq 1^\circ$, there is a
striking correlation between the morphology of the observed
$\gamma$-rays and the density of molecular clouds as traced by CS
emission \cite{CS99} (Fig. \ref{fig:GCDiffuse}~(b)). This is a strong
indication for the presence of 
a nucleonic cosmic ray accelerator in the centre of our Galaxy, since
accelerated nucleons would interact with the ambient gas of the
clouds, giving rise to the observed $\gamma$-ray flux via
$\pi^0\to\gamma\gamma$ decays. The fact that no emission is seen
further than $|l|\approx 1^\circ$ suggests that the cosmic rays
stem from a rather young source near the Galactic Centre. A simple
diffusion model assuming a source age of about $10^4$ years can well
reproduce the observed $\gamma$-ray flux distribution
\cite{GCDiffuse06}.

\subsection{Extragalactic sources} Apart from Galactic objects, \HESS\ 
observed a long list of possible extragalactic sources of VHE
$\gamma$-rays. Active Galactic Nuclei (AGN) are of particular
interest, since they are known to emit radiation in all wavebands,
including VHE $\gamma$-rays. Furthermore, these objects are known to
be highly variable. To constrain any model of AGN, data are needed for
a wide energy range, in particular at the highest energies where only a
very limited number of measurements exists.\\ \HESS\  observed a number
of known AGN and star burst galaxies. Objects like PKS 2005-489
\cite{pks2005}, Mkn421 \cite{mkn421}, or PKS 2155-304
\cite{pks2155_2004} were detected with high significance. The latter
was also observed in a simultaneous multi-wavelength campaign
covering radio, optical, X-ray, and VHE $\gamma$-ray emission, and
short-time variability of the flux has been discovered
\cite{pks2155_2005}.  For other objects, no VHE $\gamma$-ray emission
was found, and strong upper limits were obtained for 19 AGN~
\cite{AGNUL} and the star forming galaxy NGC~253 \cite{ngc253}.\\ For
very distant AGN, a significant softening of energy spectra is
expected because of the partial absorption of VHE $\gamma$-rays due to
photon-photon interactions with the Extragalactic Background Light
(EBL). For such AGN with hard observed spectra, one can therefore
constrain the amount of EBL present in the universe, which is of great
importance for 
cosmological studies. Following this logic, \HESS\  observations of the
blazars H~2356-309 (redshift $z=0.165$) and 1ES~1101-232 ($z=0.186$)
revealed that the universe is more transparent to VHE $\gamma$-rays
than previously thought \cite{EBLPaper}.

\section{The Future of \HESS\ : H.E.S.S. II} While \HESS\  continues
to take interesting data, the \HESS\  collaboration is designing and
constructing an upgrade to the existing telescope array. \HESS\  phase
II consists of a very large single telescope that will be placed in
the centre of the \HESS\  I array. The mirror collection area of \HESS\ 
II is about $600~m^2$, exceeding that of a single \HESS\  telescope by
about a factor 6. The camera has a $3^\circ$ wide field-of-view,
made up of about 2000 PMTs. The pixel size is $0.07^\circ$,
which will result in a more resolved shower image compared to
\HESS\  I.\\ 
For the operation of \HESS\  II two different modes are
foreseen: In coincidence mode, the new telescope is triggered by the
\HESS\ ~I array. In the energy range between
0.1~TeV and a couple of 10~TeV, this will increase the sensitivity of
the array by at least a factor of 2 compared to \HESS\ I. 
When operating in stand-alone mode, the energy
threshold of the new telescope will be about 30~GeV.

\section{Conclusions} The \HESS\  experiment has been fully operated
since December 2003. For the past two years it has run with full
performance and produced numerous exciting new results in VHE
$\gamma$-ray observations. \HESS\  showes for the first time that the
latest generation of ground-based Cherenkov telescope arrays has
reached a sensitivity where real $\gamma$-ray astronomy is
feasible. The sensitivity of \HESS, combined with its good angular
resolution, provides an excellent handle to study and understand the
physics processes that take place in galactic and extragalactic
sources of VHE $\gamma$-rays.\\ 
Starting in 2008, H.E.S.S. II will
explore the $\gamma$-ray sky with even higher sensitivity and at least
two-times lower energy threshold. This will provide a comfortable
overlap in energy with the upcoming GLAST \cite{GLAST} satellite
experiment.


\section*{References}
\begin{thebibliography}{99}
\bibitem{Camera} P. Vincent (for the \HESS\  collaboration),
\Journal{\em Proc. 28th ICRC (Tsukuba)}{}{281}{2003}.
\bibitem{Trigger} S. Funk {\it et al}, \Journal{\em Astroparticle
Phys.}{22}{285}{2004}.
\bibitem{Analysis} F.A. Aharonian {\it et al} (\HESS\  collaboration),
\Journal{\AA}{430}{865}{2005}.
\bibitem{Green04} D.A. Green, \Journal{\em BASI}{32}{335}{2004}.
\bibitem{Man05} R.N. Manchester {\it et al}, \Journal{\em
AJ}{129}{1993}{2005}.
\bibitem{whip04} K. Kosack {\it et al},
\Journal{\APJ}{608}{L97}{2004}.
\bibitem{Cang04} K. Tsuchiya {\it et al},
\Journal{\APJ}{606}{L115}{2004}.
\bibitem{Cang02} R. Enomoto {\it et al}, \Journal{\em
Nature}{416}{823}{2002}.
\bibitem{rxj03} F.A. Aharonian {\it et al} (H.E.S.S. collaboration),
\Journal{\em Nature}{432}{75}{2004}.
\bibitem{ScanI} F.A. Aharonian {\it et al} (H.E.S.S. collaboration),
\Journal{\SCI}{207}{1938}{2005}.
\bibitem{ScanII} F.A. Aharonian {\it et al} (H.E.S.S. collaboration),
\Journal{\APJ}{636}{777}{2006}.
\bibitem{Jon2001} F. Jones {\it et al},
\Journal{\APJ}{547}{264}{2001}.
\bibitem{MAGIC1} J. Albert {\it et al} (MAGIC collaboration),
\Journal{\APJ}{637}{41}{2006}.
\bibitem{MAGIC2} J. Albert {\it et al} (MAGIC collaboration),
astro-ph/0604197, accepted by {\em ApJ letters}.
\bibitem{rxj06} F.A. Aharonian {\it et al} (\HESS\  collaboration),
\Journal{\AA}{449}{223}{2006}.
\bibitem{GC2003} F.A. Aharonian {\it et al} (\HESS\  collaboration),
\Journal{\AA}{425}{L13}{2004}.
\bibitem{GCDiffuse06} F.A. Aharonian {\it et al} (\HESS\ 
collaboration), \Journal{\em Nature}{439}{695}{2006}.
\bibitem{pks2005} F.A. Aharonian {\it et al} (\HESS\  collaboration),
\Journal{\AA}{436}{L17}{2005}.
\bibitem{mkn421} F.A. Aharonian {\it et al} (\HESS\  collaboration),
\Journal{\AA}{437}{95}{2005}.
\bibitem{pks2155_2004} F.A. Aharonian {\it et al} (\HESS\ 
collaboration), \Journal{\AA}{430}{865}{2005}.
\bibitem{pks2155_2005} F.A. Aharonian {\it et al} (\HESS\ 
collaboration), \Journal{\AA}{442}{895}{2005}.
\bibitem{AGNUL} F.A. Aharonian {\it et al} (\HESS\  collaboration),
\Journal{\AA}{441}{465}{2005}.
\bibitem{ngc253} F.A. Aharonian {\it et al} (\HESS\  collaboration),
\Journal{\AA}{442}{177}{2005}.
\bibitem{EBLPaper} F.A. Aharonian {\it et al} (\HESS\  collaboration),
\Journal{\em Nature}{440}{1018}{2006}.
\bibitem{CS99} M.~Tsuboi {\it et al}, \Journal{{\em
Astrophys. J. Supp.}}{120}{1}{1999}.
\bibitem{GLAST} See http://glast.gsfc.nasa.gov/ and references
therein.

\end{thebibliography}
\end{document}